\documentclass[]{aa}
\usepackage{natbib}
\bibpunct{(}{)}{;}{a}{}{,} 
\usepackage{amsmath}
\usepackage{wasysym}

\usepackage{marvosym}
\usepackage{txfonts}
\usepackage{mathtools}
\usepackage{subcaption}

\usepackage{epstopdf}
\usepackage{graphicx}
\pdfoutput=1
\usepackage{hyperref}             
\hypersetup{pdfauthor=Esther Linder}
\hypersetup{backref=true, pagebackref=true, hyperindex=true, breaklinks=true,colorlinks=true,urlcolor=blue, linkcolor=blue,  citecolor=blue,pagecolor=red, bookmarks=true, bookmarksopen=true}

\def\mearth{M_\oplus}
\def\rearth{R_\oplus}

\def\f1{f_{\rm I}}

\def\beq{\begin{equation}}
\def\eeq{\end{equation}}

\def\t2{\tau_{\rm II}}

\def\sigmas0{\Sigma_{\rm s,0}}

\newcommand{\lj}{L_{\textrm{\tiny \jupiter}}}

\def\s0{S_0}

\defcitealias{marleyfortney2007}{M07}
\defcitealias{spiegelburrows2012}{SB12}

\def\apjl{ApJ}                
\def\apjs{ApJS}               
\def\ao{Appl.Optics}          
\def\aap{A\&A}                

\def\({\left(}
\def\){\right)}
\def\<{\left<}
\def\>{\right>}

\begin{document}

\title{Evolution and magnitudes of candidate Planet Nine}
\author{Esther F. Linder and Christoph Mordasini}
\institute{Physikalisches Institut, University of Bern, Sidlerstrasse 5, 3012 Bern, Switzerland} 
\date{Received 19.02.2016 / Accepted 24.03.2016}

\abstract
{The recently renewed interest in a possible additional major body in the outer solar system prompted us to study the thermodynamic evolution of such an object. We assumed that it is a smaller version of Uranus and Neptune.} 
{We modeled the temporal evolution of the radius,  temperature,  intrinsic luminosity, and the blackbody spectrum of distant ice giant planets. The aim is also to provide estimates of the magnitudes in different bands to assess whether the object might
be detectable.}
{Simulations of the cooling and contraction were conducted for ice giants with masses of 5, 10, 20, and 50 $\mearth$ that are located at 280, 700, and 1120 AU from the Sun. The core composition{, the fraction of H/He, the efficiency of energy transport, and the initial luminosity were varied.} The atmospheric opacity was set to 1, 50, and 100 times solar metallicity.}{We find for a nominal 10 $\mearth$ planet at 700 AU at the current age of the solar system an effective temperature of 47 K, much higher than the equilibrium temperature of about 10 K, a radius of 3.7 $\rearth$, and an intrinsic luminosity of 0.006 $\lj$. It has estimated apparent magnitudes of  Johnson V, R, I, L, N, Q of 21.7, 21.{4}, 2{1.0}, 20.1, 19.{9}, and 1{0.7}, and WISE W1-W4 magnitudes of 20.1, 20.{1}, 1{8.6}, and 10.{2}.  The Q and W4 band and other observations longward of about 13 $\mu$m pick up the intrinsic flux.}
{If candidate Planet 9 has a significant H/He layer and an efficient energy transport in the interior, then its luminosity is dominated by the intrinsic contribution, making it a self-luminous planet. At a likely position on its orbit near aphelion, we estimate for a mass of 5, 10, 20, and 50 $\mearth$ a V magnitude from the reflected light of 24.{3}, 23.7, 23.{3}, and 22.{6} and a Q magnitude from the intrinsic radiation of 1{4}.6, 1{1.7}, 9.{2}, and {5.8}. The latter would probably have been detected by past surveys.}


\titlerunning{Evolution of Possible Candidate Planet Nine}
\authorrunning{E.Linder \& C. Mordasini}

\maketitle

\section{Introduction}
The presence of another major body in the solar system would be of highest interest for planet formation and evolution theory. Based on the observed peculiar clustering of the orbits of trans-Neptunian objects \citep{browntrujillo2004,TrujilloSheppard2014} and after analyzing several earlier hypotheses \citep{delaFuentedelaFuente2014,iorio2014,MadiganMcCourt2016}, \citet{bat16} have recently proposed that a $\sim$10 $\mearth$ planet might be present in the outer solar system at a distance of several hundred AU from the Sun. Following the historical example of Neptune, the next step would be the observational discovery of the object. Depending on its specific properties, because of faintness, slow motion on the sky, and confusion with Galactic background stars, the object might in principle have evaded detection up to now. In this article we present predictions for the physical properties of a planet such as the one proposed, which is important to determine whether it might be detectable. Based on the most likely mass and the extrasolar planetary mass-radius relation \citep[e.g.,][]{GettelCharbonneau2016}, we assume that candidate Planet 9 has the same basic structure as Uranus and Neptune. As already speculated by  \citet{bat16}, it might be an ejected failed giant planet core. Planets of this type frequently from in planet formation simulations that are based on the core accretion theory \citep{mordasinialibert2009a} and may be scattered to large distances by giant planets \citep{BromleyKenyon2011b}.

\section{Evolution model}
Our planet evolution model \citep{mor12}  calculates the thermodynamic evolution of  planetary parameters (such as luminosity or planetary radius) over time for a wide range of initial conditions. The planet structure is simplified by assuming that the central {part} consists of iron. This is wrapped in a silicate mantel followed by a possible water ice layer and finally a H/He envelope. To
derive the luminosity, the contributions of the contraction and cooling of the gaseous envelope and of the solid core {can be} included, as well as radiogenic heating. {In the default configuration of the model, t}he interior is assumed to be adiabatic {and to fully contribute to the planet luminosity, as is probably the case for Neptune, but not for Uranus \citep{NettelmannHelled2013}. Alternatively, the solid part (core) of the planet can also be assumed to be isothermal or to be at zero temperature. In the latter case, it does not contribute to the gravothermal energy release \citep{baraffechabrier2008}}. The atmospheric model is gray and modeled with the condensate-free opacities of \citet{freedmanlustig-yaeger2014}. Stellar irradiation is included through the \citet{baraffehomeier2015} tracks. The simulations start at 10 Myr at a prespecified initial luminosity.

\section{Simulations}\label{simu}

Based on  the nominal scenario in \citet{bat16}, we started our study with a planet of 10 $\mearth$ at a semimajor axis of 700 AU. Given the typical core-to-envelope mass ratio found in planet formation simulations \citep{mor14}, we assumed the core and envelope masses to be 8.6 and 1.4 $\mearth$, respectively. This is compatible with the estimated H/He mass fractions of Uranus (12-15\%) and Neptune (16-19\%, \citealt{guillotgautier2014}). The initial luminosity was also taken from the formation simulations and set to 1.41 $\lj$, similar as in \citet{fortneyikoma2011}. The core composition is 50\% water ice (referred to as ,,ice'' in the following), 33.3\% silicates, and 16.7\% iron {(this mix of silicate and iron is} referred to as ,,rock'' in the following). The atmospheric opacity corresponds to an enrichment of 50 times solar. This first simulation is called the ,,nominal case'' in the following. To calculate magnitudes, we used a blackbody spectrum for the planet, {the Vega spectrum from \citet{bohlingilliland2004}}, and {tabulated} filter transmissions {\citep{mannvonbraun2015,besselbrett1988,persson1998,debuizer2005,jarrettcohen2011}} . We calculated the reflected flux from the object assuming it is at full phase. {In the nominal model, t}he Bond and geometric albedo were {set to}  0.31 and 0.41. These are
the values for Neptune \citep{TraubOppenheimer2010}.

For non-nominal simulations {of the 10 $\mearth$ planet, the initial luminosity was set to a value 10 times higher and lower. Such a variation covers the spread of post-formation luminosities found in our formation simulations \citep{mor14}. We also increased the H/He fraction to 50\% to explore a very efficient gas accretion during formation. The consequences of an isothermal instead of an adiabatic temperature structure in the core were studied as well. } The core composition was also varied, assuming a purely rocky and purely icy core. We also varied the opacity of the atmosphere, corresponding to an enrichment of 1 and 100 times solar, which is approximately the enrichment of carbon in Uranus and Neptune \citep{guillotgautier2014}. To account for the high eccentricity of the object, the semimajor axis was changed to 280 and 1120 AU. 

{We also studied Uranus-like cases where we set the Bond and geometric albedo to 0.30 and 0.51 \citep{TraubOppenheimer2010,guillotgautier2014}. It is well known that the current intrinsic luminosity of
Uranus is much lower than predicted by evolutionary models that assume that the interior is fully adiabatic \citep[e.g.,][]{fortneyikoma2011,NettelmannHelled2013}. As suggested by \citet{podolakhubbard1991}, this might be due to a statically stable interior that is unable to convect because of compositional gradients. If the heat is transported by conduction instead of convection, a strongly reduced flux is expected \citep{NettelmannHelled2013}. A simple way to model this situation is to exclude a central part of the planet interior from contributing to the luminosity. With this approach,  \citet{NettelmannHelled2013} found that
the low luminosity of Uranus can be matched if about half of its mass does not contribute. To investigate this possibility, we ran models where the core does not contribute to the planet luminosity, which means that we deprived the planet of its primary reservoir of gravothermal energy. An alternative explanation for the low intrinsic luminosity of Uranus might in principle be that it was already very cool when it formed, although this appears rather unlikely from formation models \citep{stevenson1982b}. To simulate such a situation, we also ran a model where the initial luminosity was chosen such that the present-day luminosity is a factor 10 lower than in the nominal case. This factor 10 approximately corresponds to the ratio of Neptune's intrinsic luminosity and to the upper limit for the intrinsic luminosity of Uranus \citep{guillotgautier2014}.} 

Finally, the planet mass was varied to 5, 20, and 50 $\mearth$ with envelope mass fractions of 10, 21, and 37 \% as  representative for the aforementioned formation simulations, from which we also took the initial luminosities of 0.32, 6.27, and 44.96 $\lj$, respectively.

\section{Results}
\subsection{Nominal simulation}\label{nominal}
\begin{figure*}[]
\begin{center}
\begin{minipage}[t]{0.48\textwidth}
\centering\includegraphics[width=0.89\linewidth]{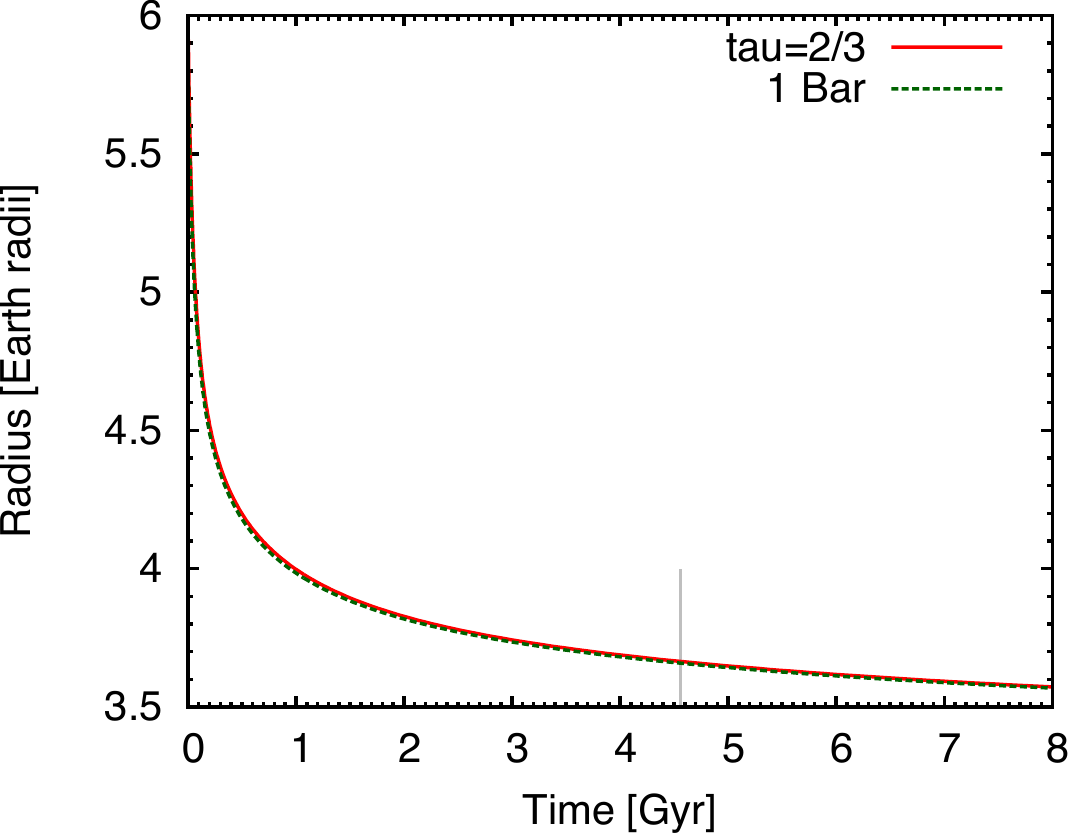}
\end{minipage}
\begin{minipage}[t]{0.48\textwidth}
\centering\includegraphics[width=0.92\linewidth]{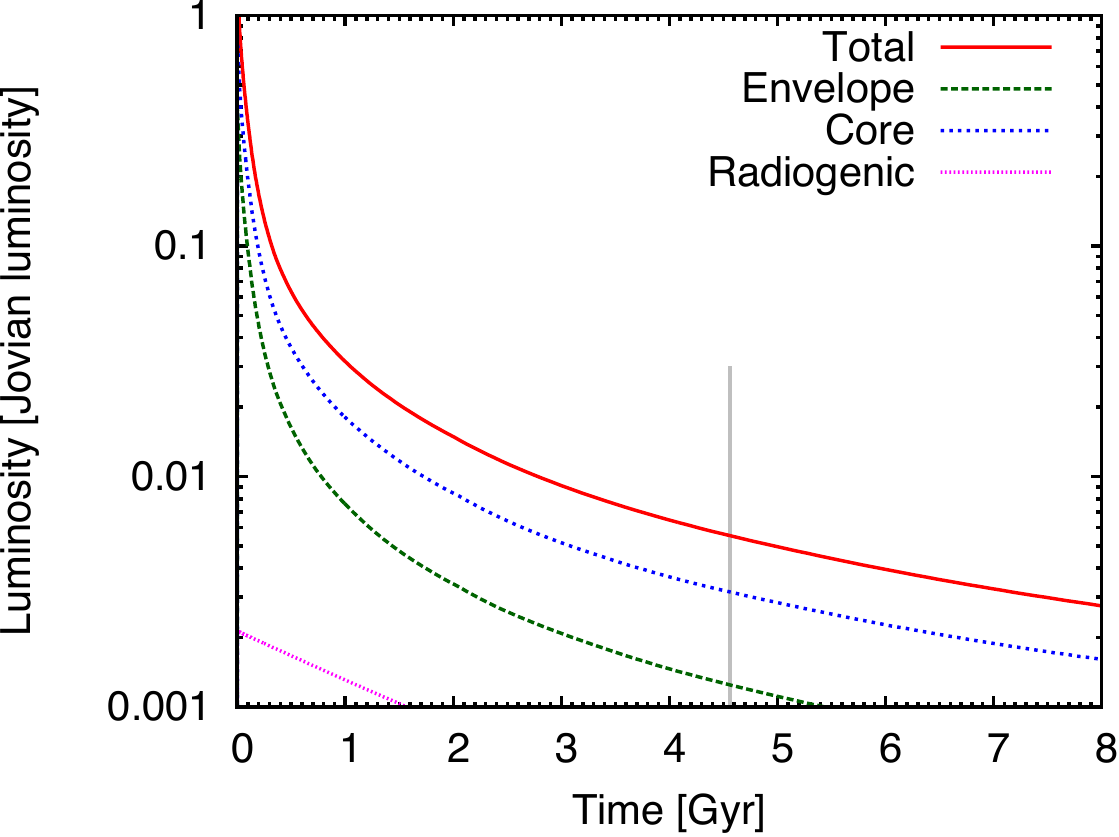}
\end{minipage}
\begin{minipage}[t]{0.48\textwidth}
\centering\includegraphics[width=0.89\linewidth]{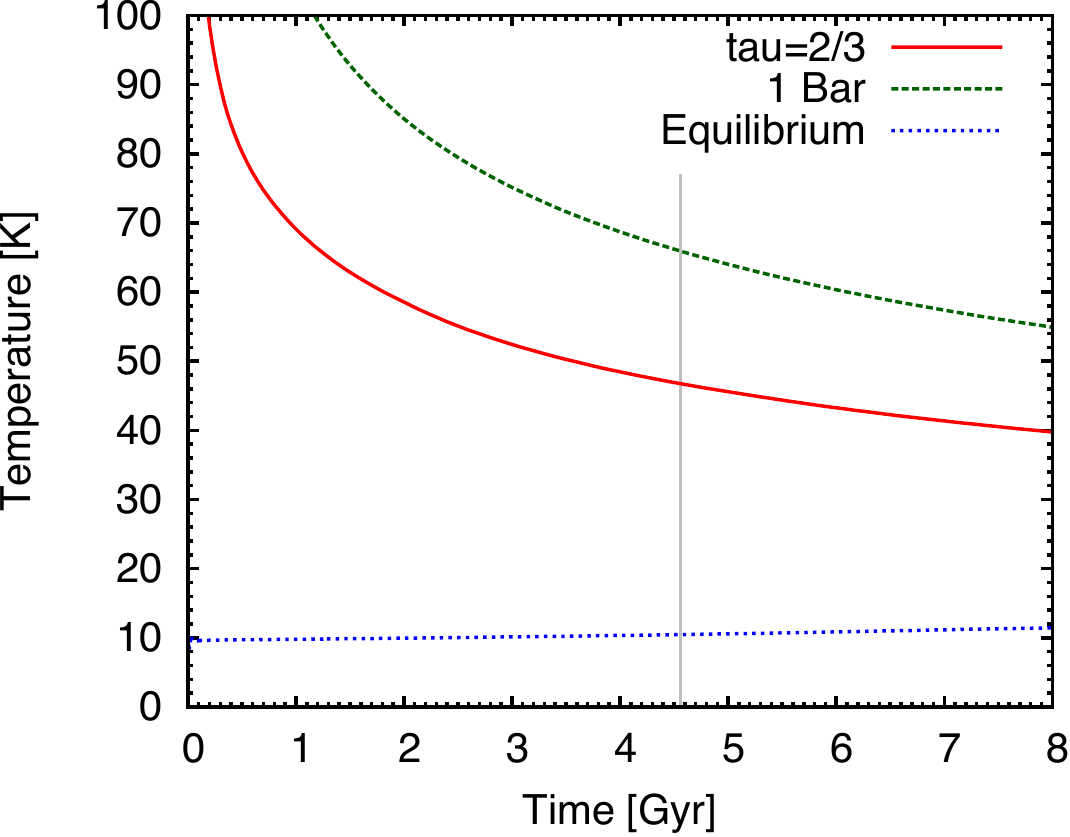}
\end{minipage}
\begin{minipage}[t]{0.48\textwidth}
\centering\includegraphics[width=1.02\linewidth]{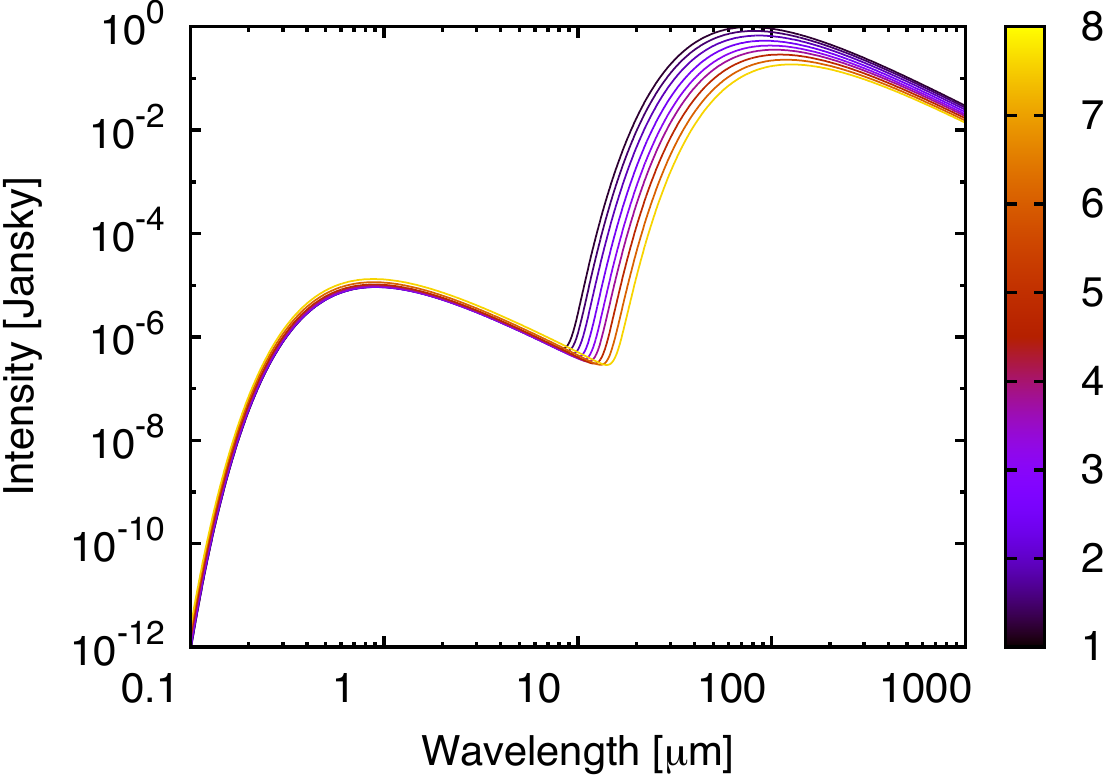}
\end{minipage}
\caption{Temporal evolution of fundamental properties of a 10 $\mearth$ planet at 700 AU with a 1.4 $\mearth$ H/He envelope, a core consisting of 50\% ice and 50\% rock, and an atmospheric
opacity of 50 times solar. \textit{Top left:} Radius. \textit{Top right:} Intrinsic luminosity. \textit{Bottom left:} Temperature. \textit{Bottom right panel:} Blackbody spectrum where the age in {gigayears} is given by the color code.}\label{nom}
\end{center}
\end{figure*}
Figure \ref{nom} shows the evolution of the radius, luminosity, temperature, and blackbody spectrum for the nominal case. The top left panel shows that the 1 bar radius today is at 3.66 $\rearth$  equal to 23'300 km. The top right panel shows the decrease of the total, envelope, core, and radiogenic  luminosity in time. The current total intrinsic luminosity is 0.006 $\lj$, with a dominant contribution from the core thermal cooling. For comparison, Neptune has an intrinsic luminosity of 0.01 $\lj$ \citep{guillotgautier2014}.  The bottom left panel shows that the current temperature at $\tau$=2/3 is 47 K. The wavelength of the highest black body emission at a temperature of 47 K follows from Wien's law and is 62 $\mu$m in the far-infrared (far-IR). This temperature is much higher than the equilibrium temperature of about 10 K at 700 AU. The planet energy budget is thus dominated by the intrinsic flux. For this, the assumption of a fully convective interior is critical. Without  efficient energy transport, the planet intrinsic luminosity could be much lower, as for Uranus \citep{NettelmannHelled2013}. {Such an Uranus-like situation is considered in Sect. \ref{variedpara}.} The temperature at the 1 bar level is about 66 K, while at the envelope-core boundary around 2100 K are reached. The equilibrium temperature also rises slowly as a result of the solar evolution. The evolution of the blackbody spectra is shown in the bottom right panel. The reflected contribution at wavelengths shorter than $\sim$13 $\mu$m remains nearly constant, whereas the higher and varying intrinsic contribution in the mid- and far-IR follows the  intrinsic luminosity. Non-gray effects can significantly modulate the actual spectrum \citep[e.g.,][]{TraubOppenheimer2010}.

\subsection{Magnitudes in various bands during one orbit}\label{kepleradded}
\begin{figure*}[]
\begin{minipage}[t]{0.49\textwidth}
\centering\includegraphics[width=1\linewidth]{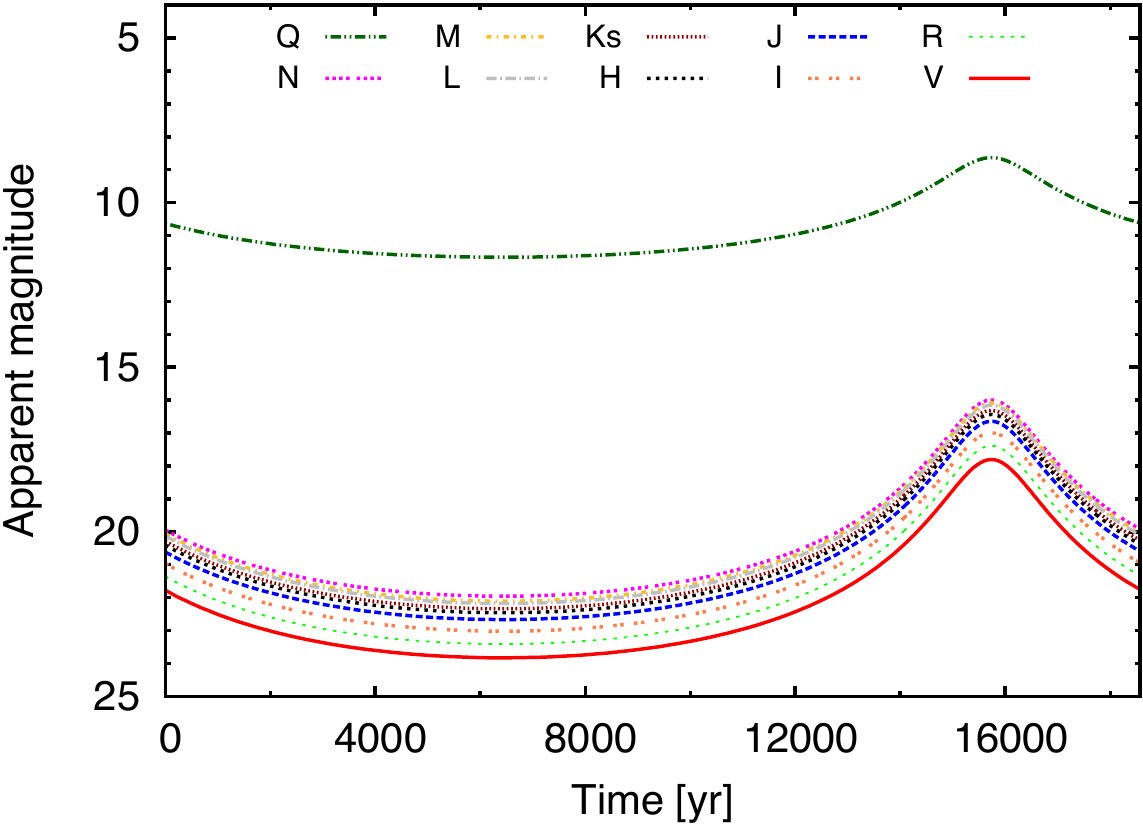}
\end{minipage}
\hfill
\begin{minipage}[t]{0.49\textwidth}
\centering\includegraphics[width=1\linewidth]{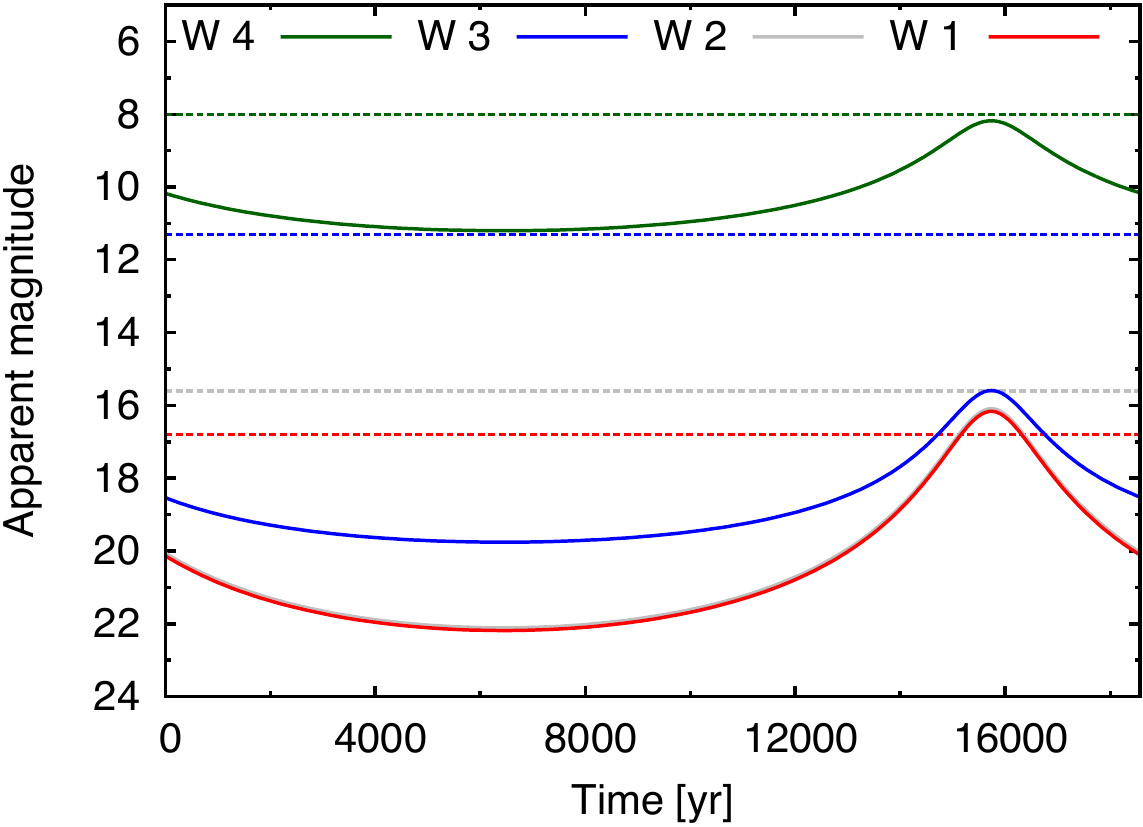}
\end{minipage}
\caption{Present-day apparent magnitudes for the nominal 10 $\mearth$ planet during one orbit around the Sun at a=700 AU and e=0.6. \textit{Left panel:} Johnson V to Q.  \textit{Right panel:} WISE filters, together with S/N=5 limits  shown by dashed horizontal lines \citep{luh14} .}\label{magfilters}
\end{figure*}

For an age of 4.56 Gyr, when the luminosity is 0.006 $\lj$ ,
we carried out a simulation with nominal parameters, for which
we varied the planet heliocentric distance in time, as found from solving the Kepler equation. The thermal timescale of the atmosphere (outer radiative zone) is of about 100 years, as found from dividing the atmospheric thermal energy content by the luminosity. This is much shorter than the orbital timescale, which means
that the atmosphere is expected to adapt to the varying stellar irradiation without significant thermal inertia. Figure \ref{magfilters} shows the apparent magnitudes for various  bands for the nominal case during one orbit around the Sun at a semimajor axis of 700 AU and an eccentricity of 0.6, as proposed by \citet{bat16}. At a distance of 700 AU, the planet has estimated apparent magnitudes of Johnson V, R, I, L, N, Q of 21.7, 21.{4}, 2{1.0}, 20.1, 19.{9}, and 1{0.7}. In the Q filter the intrinsic luminosity of the planet is visible. The planet is thus much brighter in the mid- and far-IR than in the visual or near-IR. The Wide-Field Infrared Survey Explorer (WISE) magnitudes W1, W2, W3, and W4 are shown in the right panel; they are 20.1, 20.{1}, 1{8.6}, and 10.{2} at 700 AU. The magnitudes at other heliocentric distances $r$ are simply found from the 1/$r^{4}$ dependency for the reflected flux, meaning that at perihelion it is 4 mag brighter and 2 mag dimmer at aphelion, while for the intrinsic flux (Q and W4), the $1/r^{2}$ dependency means that at perihelion it is 2 mag brighter and  1 mag dimmer at aphelion. {The W3 band picks up both reflected and intrinsic radiation.} In the right panel the WISE S/N=5 limits of W1=16.8, W2=15.6, W3=11.3, and W4=8.0 \citep{luh14} are also shown, indicating that it is only possible to see the object with WISE close to perihelion with the W1 filter. The actual detection limit of the survey of  \citet{luh14} is W2=14.5, given by the ability of detecting an object in multiple epochs. 
\subsection{Long-term evolution of the magnitudes and non-nominal simulations}\label{variedpara}
{
To understand the effect of the assumptions made for the nominal model, we simulated evolutionary tracks using different initial conditions and model settings. In the first group of simulations the planets all have a total mass of 10 $\mearth$, while in the second set of simulations we considered masses of 5, 20, and 50 $\mearth$.  }

\begin{figure*}[]
\begin{center}
\begin{minipage}[t]{0.48\textwidth}
\centering\includegraphics[width=0.99\linewidth]{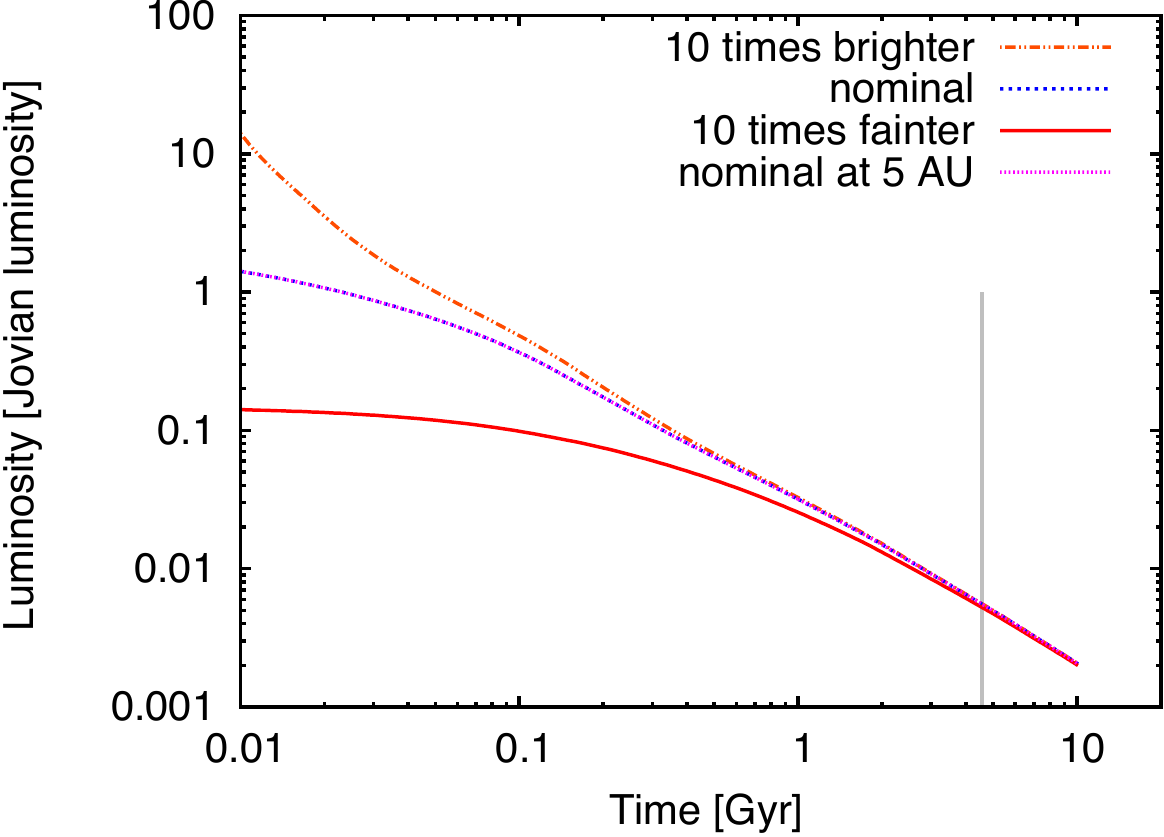}
\end{minipage}
\hfill
\begin{minipage}[t]{0.48\textwidth}
\centering\includegraphics[width=1\linewidth]{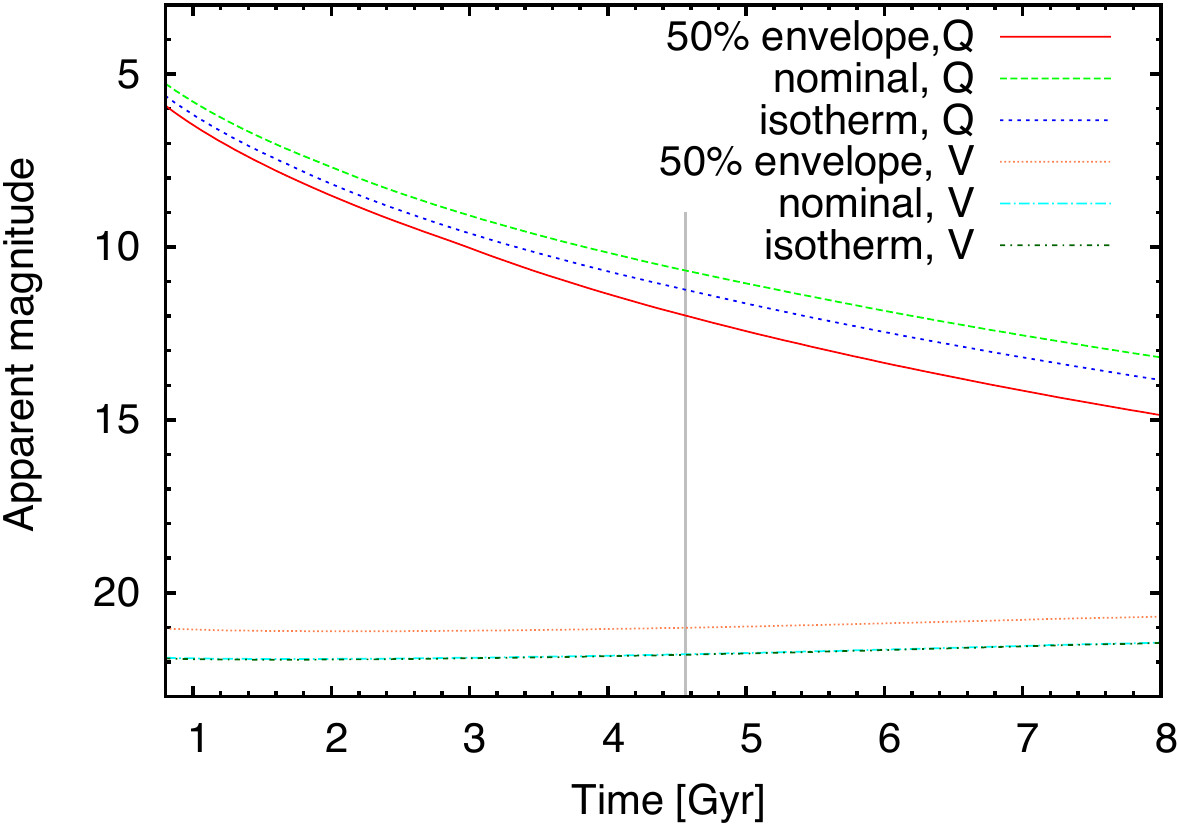}
\end{minipage}
\begin{minipage}[t]{0.48\textwidth}
\centering\includegraphics[width=1.01\linewidth]{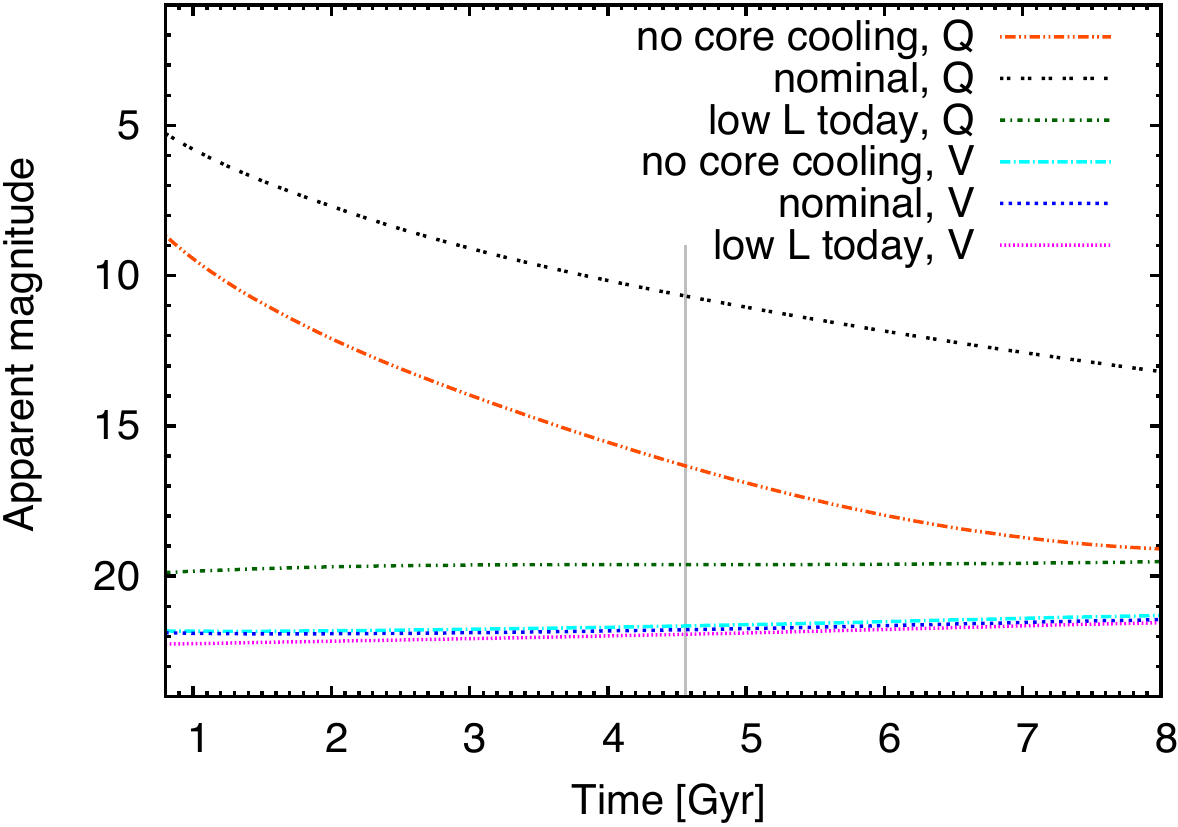}
\end{minipage}
\hfill
\begin{minipage}[t]{0.48\textwidth}
\centering\includegraphics[width=1\linewidth]{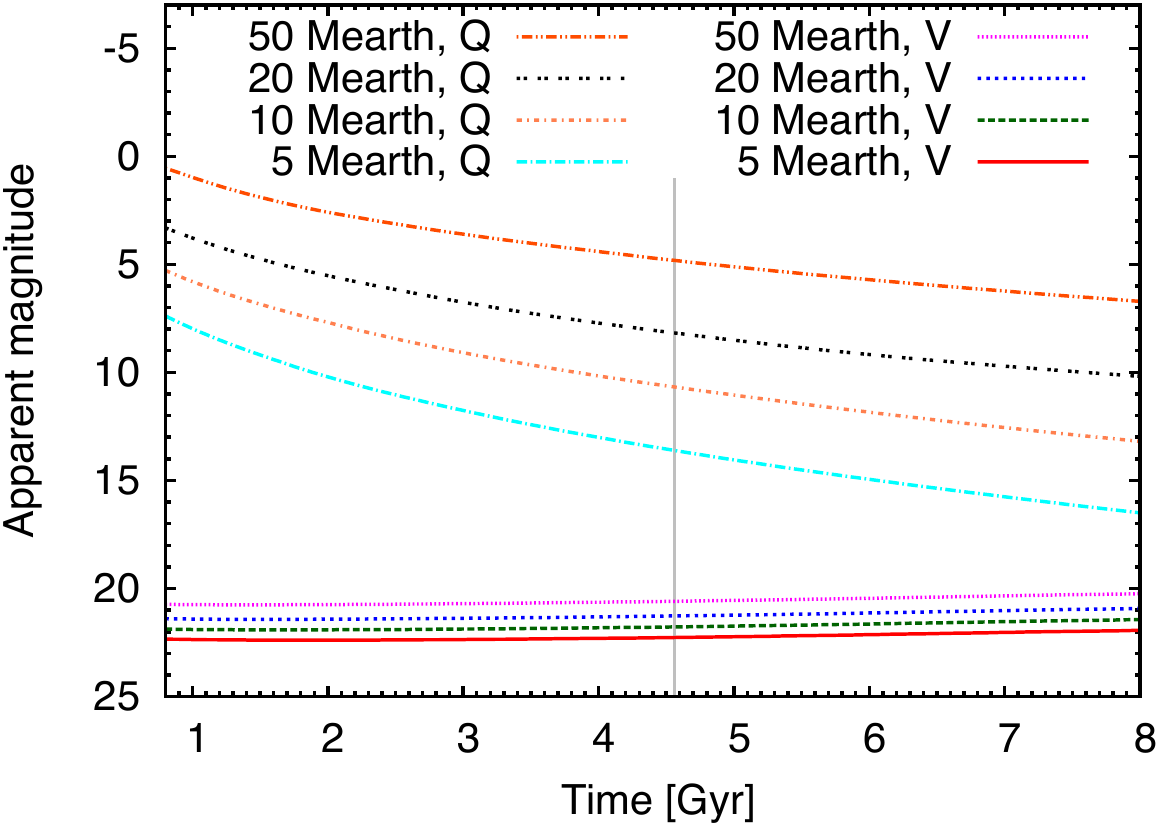}
\end{minipage}
\caption{{\textit{Top left:} Intrinsic luminosity over time for different initial conditions. \textit{Top right:} Apparent magnitudes over time for different internal structures. \textit{Bottom left:} Apparent magnitudes  for Uranus-like cases. All these three panels are for a 10 $\mearth$ planet.}  \textit{Bottom right:} Evolution of the apparent magnitudes in V and Q for the nominal case (10 $\mearth$) and planets with  masses of 5, 20, and 50 $\mearth$. All apparent magnitudes are for a distance from the Sun of 700 AU.}\label{refereefig}
\end{center}
\end{figure*}

{
\textit{Effect of different initial conditions.} In the top left panel of Fig. \ref{refereefig} we show the change in luminosity over time for different initial luminosities and also for a nominal planet but with a semimajor axis of 5 AU. This latter simulation was made to study the scenario in which the planet first evolves closer to the Sun before it is ejected to its current position. It is evident that neither the change in starting luminosity (at least for a variation by two orders of magnitude, as indicated by the formation model) nor the change in semimajor axis has a significant effect on the luminosity of the planet today. From the current thermodynamic properties of the planet  it is therefore difficult to constrain the moment in time when the planet underwent a possible scattering outward from its potential formation region near the other giant planets. 
}

{
\textit{Effect of  different internal structures.} The top right panel in Fig. \ref{refereefig} shows the temporal change in apparent Q and V magnitude for different internal structures: for a planet that has an envelope mass fraction of 50\% instead of 14\%, and for a planet whose core is isothermal instead of adiabatic. In the former case, the intrinsic luminosity increases slightly. At the same time, the magnitude in the Q band increases from 10.7 to 12.0 mag, meaning that the planet becomes dimmer in this band. This is a result of the decrease of the planet effective temperature because the increase in planet radius that is due to the higher  H/He mass fraction overcompensates for the higher luminosity. For an isothermal core, the planet heat reservoir in the core is reduced compared to the adiabatic case because the core is entirely at the temperature of the envelope-core boundary. This results in a luminosity that is lower by about 15\% than in the nominal case. This is a much weaker change than
occurs if the core contribution is entirely neglected, as we show below. We therefore recover the finding of \citet{baraffechabrier2008},
who reported that the nature of the heat transport in the core is not as important as is including the core  thermal contribution as such. In the V band, where the magnitudes range from 21.0 to 21.8, the case with a 50\% envelope mass fraction is the brightest because of its significantly larger radius of 5.22 $\rearth$, whereas the magnitudes of the nominal and isothermal case overlap because they nearly have the same radius. }

\textit{Core composition.} The influence  of a change in core composition on the magnitudes compared to the nominal case is weak. For the V  band, the planet with a rocky core is the faintest (21.9 mag), while the planet with a purely icy core is the brightest (21.6 mag), owing to a radius that is 0.3 $\rearth$ smaller and larger than in the nominal case. In the Q band the planet with the purely icy core is also brighter (1{0.2} mag) than for a purely rocky core (1{1.5} mag), now because of the higher heat capacity of water \citep{baraffechabrier2008}.

\textit{Atmospheric opacity.} A variation of the opacities corresponding to an enrichment of 1, 50, and 100 solar has a very weak effect on the V- and Q-band magnitudes for the 10 $\mearth$ planet (differences smaller than 0.{3} mag). 

\textit{Orbital distance.} We simulated the evolution of the 10 $\mearth$ planet at a fixed distance of 280 and 1120 AU (the estimated perihelion and aphelion distances of candidate Planet 9) instead of 700 AU. The resulting present-day physical properties of the planet are virtually unchanged, showing that the interior evolution proceeds as if the planet were in isolation{, as already observed for the evolution at 5 AU}. This means that the change of the planet magnitude during its orbit is simply due to the change in its heliocentric distance. For example, the V- and Q-band magnitudes at aphelion are then  2.04 and 1.02 mag fainter than at 700 AU.

{
\textit{Uranus-like cases.}
In the bottom left panel of Fig. \ref{refereefig} two simulations of 10 $\mearth$ planets with  albedos like that of Uranus are shown together with the nominal case. One simulation was conducted without core contribution to the gravothermal heat release, which
mimics stabilizing compositional gradients (,,no core cooling''). This planet has a present-day luminosity of 0.0013 $\lj$, about a quarter of the nominal case, $R$=3.48 $\rearth$, and $T_{\rm eff}$=33 K.  In the second simulation the starting luminosity was chosen such that the planet has a present-day luminosity that is a tenth of the luminosity in the nominal case (,,low L today''), mimicking a very ,,cold'' formation. The planet has a present-day radius of 3.42 $\rearth$ and an effective temperature of 27 K. The consequences for the V-band magnitude for these two cases are limited, with a variation of only 0.3 mag. Since the V band detects the reflected flux, the difference in magnitude can be explained by the small difference in radius between these cases of 0.2$\rearth$. In contrast, the effect on the Q magnitude is much
stronger, as expected. Instead of 10.7 mag as in the nominal case, Q is 16.4 for the ,,no core cooling'' and 19.6 for the ,,low L today'' case. The efficiency of heat transport in candidate Planet 9 therefore strongly influences its brightness in the IR or at mm-wavelengths \citep{cowanholder2016}, meaning that observations in this regime might be able to constrain whether candidate Planet 9 has an internal structure like Neptune or rather like Uranus \citep{cowanholder2016}. The different temporal evolution of the Q magnitude for the ,,low L today'' case is also quite clear. This can be explained by the fact that the starting luminosity is already so low that the planet barely evolves at all, and even the Q band now mirrors mostly the evolution of the Sun.
}

{
\textit{Candidate Planet 9 as a super-Earth.} As a last non-nominal model with $M$=10 $\mearth$, we also examined the possibility that candidate Planet 9 might be a super-Earth without any significant atmosphere and an Earth-like interior. We found that the planet would have a radius of about 1.9 $\rearth$. Assuming that the silicate mantel has a chondritic abundance of radionuclides \citep{mordasinialibert2012c}, we found that its radiogenic luminosity is at the current age of the solar system about $6.8\times10^{-4}\lj$ , which is remarkably only eight times lower than the total luminosity in the nominal case and about five times higher than the Earth's intrinsic luminosity \citep{gandogando2011}. For the Earth it is estimated that about half of the intrinsic luminosity is radiogenic, and the other half originates from delayed secular cooling. The effective temperature is  38 K, only 9 K lower than in the nominal case because of the smaller radius, and the apparent magnitudes in V and Q are 23.2 and 15.2. 
}

\textit{Effect of the mass.} In the bottom right panel of Fig. \ref{refereefig}{} we show apparent magnitudes in the V and Q band for the nominal planet and planets where the mass relative to the nominal case was varied. In general, the objects are brighter in the Q than in the V band. Since in the V band the reflected light of the Sun is visible, the objects become slightly brighter with time, {corresponding to} the evolution of the Sun. By contrast, the Q band shows the intrinsic radiation of the planet, and the planets slowly become fainter as a result
of cooling.  In addition to the nominal 10 $\mearth$ case, the apparent magnitude evolution is shown in the figure for planets of 5, 20, and 50 $\mearth$. These planets have a present-day radius of  2.92, 4.62, and 6.32 $\rearth$, an intrinsic luminosity of  0.0018, 0.016, and 0.078 $\lj$, and an effective temperature of {40}, 54, and 6{9} K, respectively. The V-band magnitudes are 22.{3}, 21.{3}, and 20.{6} mag, and the Q-band magnitudes are 14.6, 8.{2}, and 5.{8} mag, respectively.

\section{Discussion and conclusion}
Motivated by the recent suggestion by \citet{bat16} that there
might be a possible additional planet in the solar system, the evolution in radius, luminosity, temperature, and blackbody spectrum of a 10 $\mearth$ planet at 700 AU was studied. We assumed it to be a smaller version of Uranus and Neptune. {For an adiabatic interior,} we found that the radius of such an object today would be about 3.66 $\rearth$. The {present} intrinsic luminosity of the planet would be  about 0.006 $\lj$ with the highest contribution coming from the thermal cooling of the core. The effective temperature of the object would be much higher with 47 K than its equilibrium temperature of 10 K, meaning that the planet emission would be dominated by its internal cooling and contraction. Its intrinsic power would be about 1000 times higher than its absorbed power, making it a self-luminous planet. For comparison, Jupiter's intrinsic power is about half as high as its absorbed power \citep{guillotgautier2014}. This dominance would also be visible in the evolution of the blackbody spectrum of the planet, where the evolving intrinsic flux would be much higher than the reflected flux. This also means that the planet would be much brighter in the mid- and far-IR than in the visual and near-IR. 

The effect on the apparent magnitude of the planet in the V and Q band was studied for {different initial conditions, internal structures, different model assumptions regarding the efficiency of heat transport in the interior, and for} a variation in the planet core composition,  atmospheric opacity, semimajor axis, and mass. {For an adiabatic interior (possibly as
in Neptune, \citealt{NettelmannHelled2013})}, the strongest influence on the Q band comes from the planetary mass, whereas the strongest influence on the V band comes from the planet semimajor axis. This is expected because in the Q band, the intrinsic luminosity of the planet is visible, which is most sensitive to its mass. In contrast, the reflected light is detected in the V band, which is most sensitive to the planet distance from the Sun. {For an inefficient heat transport in the interior, possibly as in Uranus, we found that the object could be much dimmer in the Q band by 6 to 9 magnitudes, making it much harder to detect at long wavelengths. The V magnitude, in contrast, is only very mildly increased by about 0.3 mag. This means that the ratio of the brightness in the visual and in the mid- and far-IR constrains the nature of the object.} 

{We found that the evolution of the planet interior is almost identical for a semimajor axis of 5 to 1120 AU, meaning that the moment when the planet arrived at its distant position from the Sun is difficult to constrain from the current intrinsic luminosity. An early ejection during the nebular phase at ages of less than 10 Myr could be preferable from a dynamics point of view because the gaseous nebula could then enable the retention of the object on a far bound orbit \citep{bat16}. On the other hand, for such an early ejection, the planet might be lost due to stellar encounters in the solar birth cluster \citep{liadams2016}. Insights into the planet origin and formation history might be obtained from its chemical composition or spectrum.}  

Many surveys in various wavelength domains and with different sensitivities were performed in the past. For example, \citet{truj03} observed 12\% of the sky near the invariable plane in the R band with a sensitivity limit of 20.7 mag. Candidate Planet 9 is most of the time fainter than 22 mag in the R band  (Fig. \ref{magfilters}), however, and additionally, it is assumed to have an inclination of $i=30^\circ$ \citep{bat16}. This means that \cite{truj03} had only a very slight possibility of detecting candidate Planet 9. Another study was done by \citet{ell05}, who observed the sky within 6$^\circ$ of the ecliptic with a VR filter down to a limit of 22.5 mag. In this survey, the probability of detecting  candidate Planet 9 was also low because the planet orbit and the surveyed fields overlapped only slightly. The same is the case for a study by \citet{lars07}, who observed the sky in the V filter down to a magnitude of 21 within 10$^\circ$ of the ecliptic. For comparison, at aphelion at 1120 AU, we found V magnitudes of 24.{3}, 23.7, 23.{3}, and 22.{6} for masses of 5, 10, 20, and 50 $\mearth$ for candidate Planet 9. \citet{schwamb10} studied the sky within 30$^\circ$ of the ecliptic, excluding fields lower than 15$^\circ$ from the Galactic plane.  They had a relatively good coverage of the orbit of candidate Planet 9, except when it was near 13 h RA or in front of the Milky Way, which corresponds to the most likely aphelion position of candidate Planet 9 \citep{batblog}. The survey was made in the R band down to a luminosity of 21.3 mag, whereas we found R magnitudes exceeding 22 mag near aphelion even for a 50 $\mearth$ version of Planet 9. \citet{shepp11} surveyed the southern sky and Galactic plane down to -25$^\circ$  in the R filter down to a depth of 21.6 mag. The observed fields show a relatively good coverage of the orbit of candidate Planet 9 and also came closest to the requested brightness, but still did not reach it. Additionally, their astrometric precision reaches 0.5''/hr, whereas candidate Planet 9 instead has a movement of around 0.3''/hr \citep{batblog}. A search in the IR domain was performed by \citet{luh14}, who inspected the WISE data for a distant companion to the Sun (see also \citealt{SchneiderGreco2016}). The WISE satellite mapped the whole sky twice and started mapping it a third time. The detection limits of WISE together with our calculated apparent magnitudes for the nominal case were shown in Fig. \ref{magfilters}. The possibility for WISE to detect a 10 $\mearth$ Planet 9 is small because of the brightness limits. However, if the planet were more massive, a detection becomes more probable, at least from the brightness limit alone. Especially a 50 $\mearth$ mass planet would have been visible in W4 during its entire orbit, which sets an interesting upper mass limit for the planet. In summary, the current null result seems compatible with the properties
of the nominal candidate Planet 9, especially if it is at aphelion. In contrast, future telescopes such as the Large Synoptic Survey Telescope with a sensitivity down to R=26 mag \citep{luh14} or dedicated surveys should be able to find or rule out candidate Planet 9. This is an exciting perspective.

\begin{acknowledgements}
We thank M. R. Meyer, Y. Alibert, K. Heng, S. Udry, K. Zihlmann, and W. Benz for valuable inputs to the discussion. {We thank the referee Ravit Helled for a constructive review.} E.F.L. and C.M. acknowledge the support from the Swiss National Science Foundation under grant BSSGI0$\_$155816 ,,PlanetsInTime''.  Parts of this work have been carried out within the frame of the National Center for Competence in Research PlanetS supported by the SNSF.
\end{acknowledgements}

\bibliographystyle{aa} 
\bibliography{literature.bib} 

\end{document}